\documentclass[superscriptaddress,showpacs,aps,twocolumn,prb,floatfix]{revtex4-1}
\usepackage{amssymb}
\usepackage{amsmath}
\usepackage[dvips]{graphicx}
\usepackage{bm}

\newcommand{\e}{\epsilon}

\newcommand{\bea}{\begin{eqnarray}}
\newcommand{\eea}{\end{eqnarray}}

\def\beq{\begin{equation}}
\def\eeq{\end{equation}}

\begin{document}

\title{Scanning tunneling microscopy of adsorbed molecules on metalic
surfaces for nearly localized atomic states}
\author{M. Romero}
\affiliation{Instituto de Desarrollo Tecnol\'ogico para la Industria Qu\'{\i}mica
(INTEC-CONICET-UNL) G\"uemes 3450, CC91, (S3000GLN), Santa F\'e, Argentina}
\author{A. A. Aligia}
\affiliation{Centro At\'omico Bariloche and Instituto Balseiro, Comisi\'on Nacional de
Energ\'{\i}a At\'omica, 8400 Bariloche, Argentina}
\date{\today}

\begin{abstract}
We consider a Hubbard-Anderson model which describes localized orbitals in
five different sites hybridized both among themselves and with a continuum
of extended states. A square planar geometry with an atom at the center is used to
represent TBrPP-Co molecules. When the renormalized effective hopping between sites is small
compared with a Kondo energy scale determined by the site-continuum hybridization, 
the system can be described
as a set of independent Kondo resonances, rather than molecular states. We
study the crossover between both regimes and analyze the spectral density of
conduction electrons as a function of position. The
results are in qualitative agreement with measurements of the differential conductance 
in a system with TBrPP-Co molecules adsorbed on a Cu(111) surface.
\end{abstract}

\maketitle


\section{Introduction}

The transport of electrons through molecules is relevant in many branches of
science and important for potential applications in electronic devices \cite%
{ven,wang,wwang} and molecular spintronics \cite{iancu,pere}. On the other
hand, the study of many-body phenomena in nanoscale systems has attracted
much attention in recent years. In particular the Kondo effect, which arises
when a magnetic impurity interacts with a continuum of extended states, has
been observed in a variety of nanoscopic systems \cite%
{iancu,pere,mad,man,knorr,jan,jari}. Scanning tunneling spectroscopy (STS)
has made it possible to probe the local density of states and Fano
antiresonances have been observed for several magnetic systems on metal
surfaces \cite{iancu,pere,mad,man,knorr,jan}. These antiresonances observed
in the differential conductance, reflect a dip in the spectral density of
conduction states near the Fermi level caused by the Kondo effect \cite{revi}%
. The half width of the dip is given by a characteristic energy scale, the
Kondo temperature $T_K$. Furthermore, corrals built on the (111) surface of
noble metals or Cu have been used to project the spectral features of the
Fano-Kondo antiresonance (FKA) to remote places \cite{man,revi}. The
observed Fano line shapes for one magnetic impurity on these surfaces have
been reproduced by many-body calculations \cite{revi,uj,mir,mir2,meri,lin,morr10}.
For corrals, predictions of the variation of $T_K$ with the position of the
impurity were made \cite{revi,morr6}.

When several magnetic impurities interacting between them are present, the
theoretical analysis is more difficult. For Cr trimers on the Au(111)
surface, controversial results were obtained \cite{kuda,sav,laza,inge} which
were in contradiction with at least part of the puzzling observed dependence
of $T_K$ with geometry \cite{jan}. More recently, a qualitative explanation
was provided
using a Hubbard-Anderson model (HAM) described in Section II \cite{tri}.

Recent STS results for TBrPP-Co [5, 10, 15,
20-Tetrakis-(4-bromophenyl)-porphyrin-cobalt] molecules adsorbed on a
Cu(111) surface are also surprising \cite{pere}. The width of the FKA
depends on the \emph{position} of the microscope tip along the molecule,
varying within a factor two among different positions \cite{note}. Quite
generally, for any system in which only one atom is highly correlated (say
Co), if the spectral density of the system without this magnetic impurity is
featureless around the Fermi energy $\epsilon_F$, the STS near $\epsilon_F$
is dominated by the Green's function of the impurity and the width of the
FKA is the same everywhere. Only the intensity varies with position (Eqs.
(11) to (14) of Ref. \cite{revi}). 
Experimentally, for a sysyem with one impurity at the focus of an elliptical quantum corral, the observed FKA is similar at both foci and only the amplitude differs \cite{man}
In systems with several magnetic atoms
with a sizable hopping between them, one can define an effective Hamiltonian
in terms of \emph{molecular} extended orbitals and if Kondo physics is
present, again the energy dependence of STS near $\epsilon_F$ is essentially
independent of position \cite{kuda,sav,laza,inge,tri,soc}. This results are
confirmed by our calculations on the HAM. 

In Ref. \cite{pere}, the
experimental results were interpreted in terms of a \emph{single} impurity
Kondo model, with a position dependent exchange interaction which increases
linearly with the local spin density. This is strictly valid only in the
limit in which the system can be considered as a continuum of independent
magnetic impurities. This situation resembles the Kondo lattice in the limit
in which $T_K$ is larger than the effective intersite exchange $I$ \cite%
{doni}. However, in the molecular system, the competing parameter is the
magnitude of the intersite hoppings $t$ (rather than $I$), neglected in the
simplified analysis of Ref. \cite{pere}. Thus, one expects a crossover
between a regime with independent 
``impurities''
for small $t$ to a collective behavior for
large $t$. In the limit of an infinite molecule disconnected to the metal,
one expects that this crossover turns to a Mott localization transition \cite{xyz}.

In this work, we describe the system with a HAM in which each magnetic
site is hybridized with a continuum of extended states and in addition there
is a hopping between magnetic sites. We obtain that for reasonable parameters (most of them used
before), the essential features of the experiment of Ref. \cite{pere} are
reproduced. Moreover we study the above mentioned crossover between
localized and extended behavior.

In Section II, we describe the model. The STS spectra, comparison with experiment, and the crossover from the regime of nearly isolated impurities to clllective molecular behavior (either with one Koddo peak or with split Kondo peaks) is presented in Section III. Section IV contains a short summary and discussion. 

\section{Model, approximations and parameters}

The Hamiltonian can be written as \cite{tri} 
\begin{eqnarray}
H &=&H_{\text{mol}}+H_{\text{met}}+H_{V}\text{,}  \notag \\
H_{\text{mol}} &=&\sum_{i\sigma }E_{i}d_{i\sigma }^{\dagger }d_{i\sigma
}+U_{i}d_{i\uparrow }^{\dagger }d_{i\uparrow }d_{i\downarrow }^{\dagger
}d_{i\downarrow }  \notag \\
&&-\sum_{\langle ij\rangle \sigma }t_{ij}(d_{i\sigma }^{\dagger }d_{j\sigma
}+\text{H.c.}),  \notag \\
H_{\text{met}} &=&\sum_{\mathbf{k}\sigma }\epsilon _{\mathbf{k}}c_{\mathbf{k}%
\sigma }^{\dagger }c_{\mathbf{k}\sigma },  \notag \\
H_{V} &=&\sum_{\mathbf{k}j\sigma }(V_{j}e^{i\mathbf{k\cdot R}_{j}}d_{j\sigma
}^{\dagger }c_{\mathbf{k}\sigma }+\text{H.c.}).  \label{ham}
\end{eqnarray}%
$H_{\text{mol}}$ is a Hubbard model that describes the isolated molecule. Here
$d_{i\sigma }^{\dagger }$
creates an electron at site $i$ lying at $\mathbf{R}_{i}$ with spin $\sigma $. 
$H_{\text{met}}$ corresponds to the conduction band of the metallic substrate, and $H_{V}$
is the hybridization between the atoms of the molecule and the extended
states of the metal, assuming that the wave functions of the latter are
plane waves \cite{revi}. Following earlier work \cite{chi,agui} we represent
the molecule with 5 sites, one at the center representing the Co atom and
four at the corners of a square 
representing bromophenyl lobes. The model has $C_{4v}$ symmetry.
We assume that the tip of the scanning tunneling microscope (STM) is far enough
from the system in such a way that it does not alter its electronic structure.

This model is similar to the one proposed by Chiappe and Verges for similar
molecules \cite{chi,agui}. The main difference is that we consider that not
only Co, but all sites are highly correlated, as suggested by recent 
\textit{ab initio} calculations \cite{pere}. In addition $H_{V}$ is somewhat more
realistic and contains phase factors $\exp (i\mathbf{k}\cdot \mathbf{R}_{j})$, 
which were essential in the description of Cr trimers on Au(111) \cite{tri}.
We also take $U_{i}\rightarrow +\infty $. 
Since the $U_{i}$ are expected to be of several eV, the $t_{ij}$ 
were proposed of the order of 0.2 eV, and the resonant level widths $\pi \rho V_j^2$ are smaller than 1 eV (see below), 
this assumption is not a serious one, and on the other hand allows us
to use the slave-boson mean-field approximation (SBMFA) introduced by Coleman \cite{soc,boso,agu}.

In this treatment, the creation operators for localized electrons are represented 
as a product of a boson and a new fermion as
$d_{i\sigma }^{\dagger }=b_i f_{i\sigma }^{\dagger }$, with the constraint
$b_i^{\dagger }b_i+ \sum_\sigma f_{i\sigma }^{\dagger }f_{i\sigma }=1$. The 
approximation consists in replacing  
the boson operators $b_i^{\dagger }$, $b_i$ are replaced by 
their expectation values $\langle b_i^{\dagger } \rangle = \langle b_i \rangle^*$
obtained minimizing the free energy.   
This approximation reproduces correctly the spectral density
near the Fermi level, and the exponential dependence 
of $T_{K}$ with the parameters in the Kondo regime. 

As in a previous work \cite{tri}, we have approximated the angular average
of  $\sum_{\mathbf{k}}\exp [i\mathbf{k}\cdot (\mathbf{R}_{i}-\mathbf{R}_{j})]
$, by its value at the Fermi surface $\beta _{ij}$. For an isotropic
three-dimensional (3D)  band $\beta _{ij}=\sin (k_{F}r_{ij})/(k_{F}r_{ij})$,
where $k_{F}$ is the Fermi wave vector and  $r_{ij}=|\mathbf{R}_{i}-\mathbf{R}_{j}|$. 
We take $k_{F}R=3.8$, where $R$ is the distance between the Co site
at the center and any lobe at the corners \cite{tri}. We also assume for simplicity a
constant density of conduction states $\rho =0.1/$eV. The results practically do not change
for other values of $\rho$ if the hybridizations $V_j$ are scaled keeping
$\rho V_j^2$ constant. 
Following Ref. \cite{agui}, 
we take $t_{0}=0.15$ eV for the Co-lobe hopping and $t_{1}=0.2$ eV
for the lobe-lobe one. From experiment \cite{iancu}, the first occupied
orbital is located 0.7 eV below the Fermi level $\epsilon _{F}$ and is assigned to the 
3d$_{z^{2}}$ orbital of the Co atom. Choosing the origin of energies at $\epsilon _{F}=0$, 
then $E_{\text{Co}}=-0.7$ eV and the on-site energy for the lobes 
should be smaller. We take  $E_{l}=-3$ eV. 
The hybridizations $V_{\text{Co}}=0.68$ eV and $V_{l}=1.68$ eV
were chosen to reproduce approximately the observed width of the resonances when the
tip is on a specific site.

\section{Results for the tunneling spectra}

The resulting values of the expectation values of the boson operators 
$\langle b_{\text{Co}} \rangle=0.18$ and $\langle b_l \rangle=0.13$ 
indicate a strong renormalization of the quasiparticles at the Fermi energy.
In particular, the effective $t_{0}$ is renormalized to 
$t_{0}\langle b_{\text{Co}} \rangle\langle b_{\text{Co}} \rangle=3.6$ meV.
This is smaller than the reported $T_K$. This means that the system is in fact 
in a regime of rather independent ``impurities'' with two different $T_K$.
While this picture is useful to gain physical insight, the intersite hopping
and the iteractions between sites through the conduction band are still important 
and affect the width and shape of the STS features, as explained below.
  
The fact that the renormalization is stronger for the lobes 
($\langle b_l \rangle < \langle b_{\text{Co}} \rangle$) is consistent with  
results of \textit{ab initio} calculations which find a larger spin density 
at the bromophenyl units \cite{pere}. 

\begin{figure}[tbp]
\includegraphics[width=8cm]{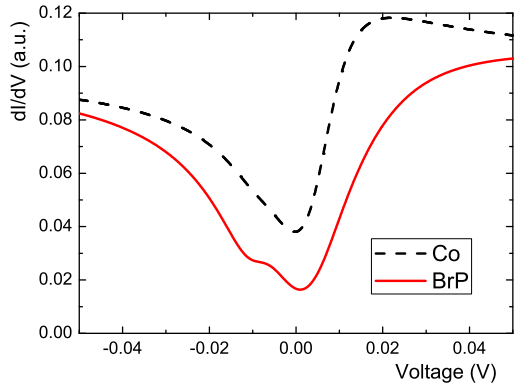}
\caption{(Color online) Differential conductance as a function of voltage when
the tip is on Co (dashed line) or on a lobe (full line)
for $q_{\text{Co}}=0.136$, $q_l=0.114$.}
\label{g1}
\end{figure}

For weak coupling of the STM tip to the system, the measured differential conductance $dI/dV$
is proportional to the density of states of a mixed operator $h_{\sigma }$ which depends on the position 
$\mathbf{R}$ of the tip. For a voltage difference $V$ between the metal and the tip one has \cite{revi,morr10}

\bea
 dI/dV &\sim& \rho_h (eV), \nonumber \\
\rho_h (\omega) &=& \frac{i}{2 \pi} \left( \langle \langle h_{\sigma };
h_{\sigma }^{\dagger }\rangle \rangle _{\omega +i\epsilon }-\langle \langle h_{\sigma };
h_{\sigma }^{\dagger }\rangle \rangle _{\omega -i\epsilon }\right) , \nonumber \\
h_{\sigma }(\mathbf{R}) &=& \sum_{\mathbf{k}}e^{i\mathbf{k\cdot R}_{j}}c_{\mathbf{k}\sigma
}+\sum_j q_{j}(\mathbf{R}) d_{j\sigma },
\label{didv}
\eea  
where $q_{j}(\mathbf{R})$ is proportional to the hopping between
the tip and the site  $j$ and controls the line shape and its asymmetry 
\cite{revi,morr10}. It is only significant when the tip is right above the site $j$. 
Then, here we assume $q_{j}(\mathbf{R})=q_{j} \delta (\mathbf{R}^{||}-\mathbf{R_j}^{||})$, 
where the superscript on the position vectors denote the component parallel to the 
Cu(111) surface.

The resulting $dI/dV$ when the tip is either on Co or on the
lobes  is shown in Fig. \ref{g1}. The values of $q_{j}$  
were chosen to result in a
line shape that agrees with experiment \cite{pere}. In any case, these
values are small indicating that  $dI/dV$ is dominated by the conduction
density of states, which presents a Fano-Kondo antiresonance, 
as for Co impurities on Cu or noble metal surfaces \cite{revi}. 
A fit of the Kondo temperature using the simple expression given in
Ref. \cite{knorr} leads to $T_{K}=112$ K on Co and $T_{K}=208$ K on the lobes.
The main features of the experiment are reproduced. When the tip is on the
lobes, we obtain a structure near -10 meV. While the experimental resuls seem rather noisy,
a similar structure seems to be present only in some of the observed spectra 
(for example $T$= 109 K, 185 K, 212 K in Fig. 3 (b) of Ref. 5). In any case, 
this structure disappears if a broadening of the quasiparticles at finite energies (not taken into
account in the SBMFA) \cite{wol}, or  slightly smaller $t_{ij}$ or larger $V_{i}$ are used.

\begin{figure}[tbp]
\includegraphics[width=8cm]{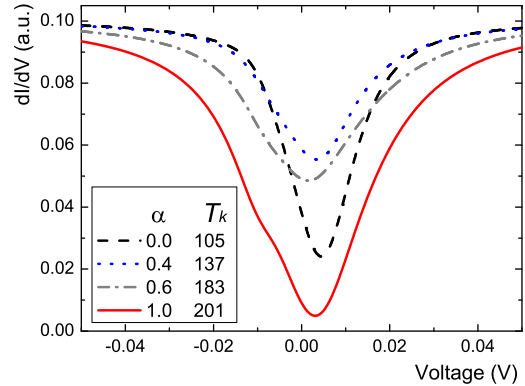}
\caption{(Color online) Differential conductance as a function of voltage for $q_j=0$ and
different positions of the tip along the Co-lobe path.}
\label{g2}
\end{figure}

In Fig. \ref{g2} we show the local conduction density of
states $\rho_h(eV)$ for $q_{j}=0$ (proportional to $dI/dV$) 
as a function of position along a path between Co and a lobe 
$\mathbf{R=\alpha R}_{l}+(1-\mathbf{\alpha )R}_{\text{Co}}$. Starting at the
Co site ($\alpha =0$), the dip broadens and loses intensity. The intensity
is recovered at the position of the lobe while the peak continues to
broaden. The space dependence of the depth of the structure might be expected 
from what we know from the
one-impurity case, where the amplitude of the dip decays as $1/d^{2}$ ($1/d$) 
with the distance to the impurity $d$ for a 3D (2D) conduction band \cite{knorr}.  
Therefore, the amplitude of the dip has a smaller variation if a
2D band or smaller $k_{F}$ is assumed. In particular, for $k_F =0$,
the dependence of the conduction density of states near $\e_F$ with position disappears.

\begin{figure}[tbp]
\includegraphics[width=8cm]{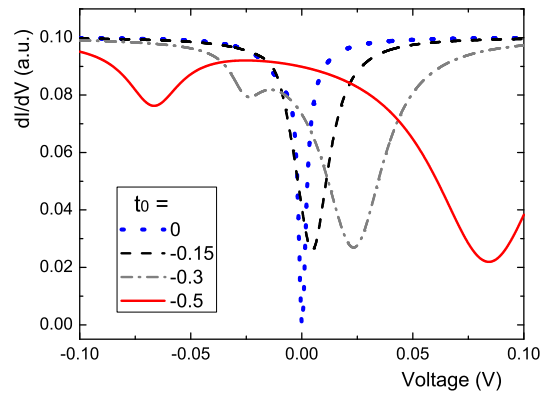}
\caption{(Color online) Differential conductance as a function of voltage for $q_j=0$ and
different Co-lobe hopping $t_0$ with the tip on top of the Co atom.}
\label{g3}
\end{figure}

To study the crossover to the regime in which molecular states dominate the physics,
we have increased the Co-lobe hopping $t_0$. The result for $dI/dV$ when the tip
is on top of the Co atom is shown in Fig. \ref{g3}. 
For $t_0=0$ a fit of the curve leads to $T_K=33$ K. This agrees with the fact that 
for an isolated Co atom on a Cu(111) surface, the observed $T_K$ (Ref. \cite{knorr})  
is about a factor 1/2 of that in the molecule (in our case however, there is a remaining 
interaction with the lobes through the bulk states unless $k_F \rightarrow \infty$). 
As $t_0$ increases, the peak near the Fermi energy 
splits in two for large enough $t_0 \sim 0.2$ eV, and the splitting increases
with $t_0$. 
The physics of this splitting is also present in previous experimental \cite{oos,bli} and
theoretical \cite{agu,agu2,zit,dias,vau2} studies of two quantum dots, and 
for impurities with hopping to discrete extended states (as surface states inside a corral with 
hard walls \cite{mir,mir2}). 
This structure is also apparent when the tip is on the lobes (not shown), 
but an additional structure at $\e_F$ remains due to states that do not mix with
the Co ones for symmetry reasons. This structure tends to split when the molecule is 
distorted to a rectangular shape. However, the effect of this distortion is not dramatic if the other parameters 
are kept constant. Nevertheless, it is known that this distortion alters the Co-surface distance and
as a consequence $V_{\text{Co}}$ should be modified \cite{iancu}.

For large $t_0$ and positive $E_l$, we have verified that there is a Kondo effect with 
a molecular state, and $dI/dV$ has a similar shape near the Fermi level for any position of the tip, 
with the same $T_K$ within less than 1\%. 
This situation is analogous to the physics of quantum corrals \cite{man,revi,mir,mir2} and 
Cr trimers on Au(111) \cite{jan,kuda,sav,laza,inge,tri}.

\begin{figure}[tbp]
\includegraphics[width=8cm]{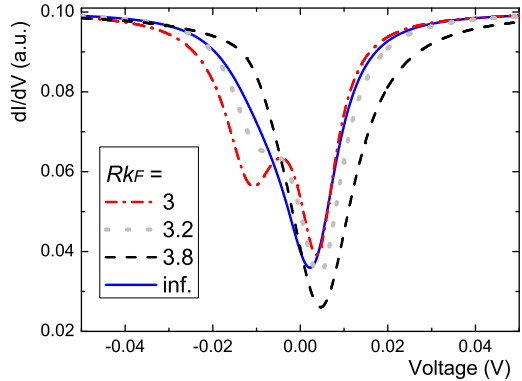}
\caption{(Color online) Differential conductance as a function of voltage for $q_j=0$ and
different Fermi wave vectors with the tip on top of the Co atom.}
\label{g4}
\end{figure}

The value of $k_F$ affects the interference effects mediated by the conduction band, 
between different sites of the molecule. For $k_F \rightarrow 0$, all conduction states near 
the Fermi energy mix with
the same molecular orbital regardless of its direction resulting in a maximum interference,
For $k_F \rightarrow \infty$ the interference occurs only through the hoppings $t_j$. 
In Fig. \ref{g4} we show how the conduction electron density of states at the Co
site changes with $k_F$. For the smaller values investigated, there is a splitting of the 
Fano-Kondo antiresonance 
due to an effective mixing of the molecular states mediated by the band.

\section{Summary and discussion}

For a molecule 
with strongly correlated states near the Fermi energy on top of a metallic surface, 
we have studied the competition between interatomic hopping 
and the hopping of these atoms
with the continuum of extended metallic states. 
We show that strong correlations can 
lead to the rather unexpected situation in which the hopping to the metallic substrate dominates, 
leading to a physics 
characterized by different Kondo impurities as a first approximation. 
The resulting $dI/dV$ and its dependence with position agrees with recent
measurements on TBrPP-Co molecules adsorbed on a Cu(111) surface.  
 We have also analyzed the crossover to the most usual regime in which molecular extended states dominate the physics. Our approximation (slave bosons in the mean-field approximation) is known to describe qualitatively the physics near the Fermi level in both limits \cite{revi,soc,agu,vau2}

\section*{Acknowledgments}

We thank CONICET from Argentina for financial support. This work was
sponsored by PIP 11220080101821 of CONICET and PICT R1776 of the
ANPCyT.

\end{document}